\documentclass[aps,preprint,floatfix,nofootinbib,showpacs]{revtex4-1}
\usepackage{graphicx,color,epstopdf,amsmath}
\newcommand{\beq}{\begin{equation}}
\newcommand{\eeq}{\end{equation}}
\newcommand{\be}{\begin{eqnarray}}
\newcommand{\ee}{\end{eqnarray}}

\newcommand{\gcs}{g_{cs}}
\newcommand{\zsc}{{Z'_{cs}}}

\newcommand{\gev}{{\,{\rm GeV}}}
\newcommand{\tev}{{\,{\rm TeV}}}
\newcommand{\gm}{{\gamma}}

\newcommand{\pb}{{\bar{p}}}
\begin{document}


\title{
The CDF dijet excess and \\
$\zsc$ coupled
to the second generation quarks
}

\author{Sanghyeon Chang}
\email{sang.chang@gmail.com}

\author{Kang Young Lee}
\email{kylee14214@gmail.com}

\author{Jeonghyeon Song}
\email{jhsong@konkuk.ac.kr}

\affiliation{
Division of Quantum Phases \& Devices, School of Physics, 
Konkuk University, Seoul 143-701, Korea
}

\date{\today}

\begin{abstract}
Recently the CDF collaboration has reported
the excess 
in the dijet invariant-mass distribution
of the $Wjj$ events, 
corresponding to a significance of 3.2 standard deviations.
Considering  the lack of similar excesses in the
$\gm jj$ and $Z jj$ events yet,
we propose a new $\zsc$ model: $\zsc$
couples only to the second generation quarks.
Single production of $\zsc$ as well as 
associated production with $W,\gm, Z$
are mainly from the sea quarks.
Only $W \zsc$ production has additional contribution
from one valence quark and one sea quark, which is
allowed by CKM mixing.
We found that if the new gauge coupling is large enough,
marginally permitted by perturbativity,
this new model can explain the observed CDF $Wjj$ anomaly
as well as the lack of $\gm jj$ and $Z jj$ anomalies.
Vanishing coupling of $\zsc$-$b$-$\bar{b}$
protects this model from the constraint
of $p\pb\to WH \to \ell\nu b \bar{b}$.
\end{abstract}

\maketitle

Recently the CDF collaboration has reported 
the excess in
the dijet invariant-mass distribution 
associated with a $W$ boson,
using 4.3 fb$^{-1}$ data at the Tevatron \cite{cdf}.
Such a disagreement with the standard model (SM)
prediction to the 3.2 $\sigma$ significance 
can be accommodated by adding a resonant excess
in the $120-160$ GeV mass region, 
which suggests a possibility of a new particle beyond the SM.
According to the CDF fit result,
the Gaussian peak of the resonance is around 145 GeV.
The number of excess events are $156 \pm 42$ for the electron
and $97 \pm 38$ for the muon, where the leptons come from $W$ decays.
With the number of observed excess,
the production cross section 
multiplied by the particle branching ratio
into dijets is estimated to be of the order of 4 pb.
A few new physics models beyond the SM 
have been suggested to explain this CDF $Wjj$ anomaly,
including leptophobic $Z'$ boson \cite{kingman,light,zp1,Anchordoqui},
supersymmetry with R-parity violation \cite{susy}, technicolor \cite{techni}, color octet boson \cite{color},
a flavor symmetry \cite{flavor} and  nucleon intrinsic quark \cite{intrinsic}.  
Among these, 
models with leptophobic $Z'$ boson 
\cite{kingman,light,zp1,Anchordoqui} 
are the most explored in a short time period.
There are some claims that this anomaly can be
explained within the SM \cite{SM:explanation}. 

In order to probe the nature of the new physics,
we have to consider more phenomenological hints 
in addition to the CDF $Wjj$ anomaly.
First, it is not likely that the resonant excess is due to
the Higgs boson production in and beyond the SM.
The estimated $\sigma \times Br(jj)$ of the excess 
is much larger than
that of the $WH$ production of 12 fb in the SM for $m_H=150$ GeV.
Moreover the CDF searches 
for $\ell \nu b \bar{b}$ final states have found
no significant excess in this mass region.
Second, 
the light $Z'$ model is to be leptophobic.
The LEP and Tevatron dilepton searches provided 
a very strong constraint on the $Z'$ boson mass.
Third, the single production of this 
leptophobic $Z'$ is constrained to some extent,
which
is subject to
the two jet searches at the Tevatron \cite{Tevatron_dijet}
and UA2 experiment at $Sp\bar{p}S$ \cite{UA2}.
In the mass region of $m_{Z'} < 200$ GeV, 
the Tevatron two jet searches are not relevant
because of overwhelming QCD background.
The UA2 constraints are most important.
To evade the UA2 constraints, couplings of $Z'$ should be 
$g_{uuZ'} < 0.4$ and 
$g_{ddZ'} < 0.4$ for $m_{Z'} \sim 150$ GeV \cite{light}.

The final hint comes from other associated productions of
$Z jj$ and $\gamma jj$. 
Most leptophobic $Z'$ models assume the universal couplings
with all generation quarks.
Similar excesses in the dijet invariant-mass ($M_{jj}$) distribution 
of the $Z jj$ and $\gamma jj$ events are expected as in the 
$W jj$ channel.
Due to the smallness of couplings,
$Z jj$ and $\gamma jj$ production cross sections are smaller
than that of $W jj$.
Moreover small leptonic branching ratio of $Z$ boson 
suppresses the signal more
as we identify the $Z$ boson by leptonic decays:
it is not promising to find the excess at the $ZZ' \to Z jj$ channel.
Instead  the $\gamma Z' \to \gamma jj$ signal at the Tevatron
should be observed if the $Wjj$ anomaly is explained 
solely by the universal leptophobic $Z'$ model
\cite{kingman}. 
At present, however, no significant indication of new physics
has been found in $\gamma jj$ channel
using 4.8 fb$^{-1}$ data of CDF at the Tevatron
\cite{gamjj}.
It might be a shortcoming of the universal 
leptophobic $Z'$ explanation 
for the CDF excess.

Summarizing the phenomenological signatures, we need a new
resonance to explain the following features:
(\textit{i}) the excess of $Wjj$ events in the dijet invariant-mass
range of $120-160\gev$;
(\textit{ii}) suppressed single production of the new resonance;
(\textit{iii}) very suppressed decays into leptons;
(\textit{iv}) no significant excess in the $\ell \nu b\bar{b}$ mode
at the Tevatron;
(\textit{v}) not statistically significant excess of $\gm jj$ events.

\begin{figure}[t!]
\centering
\includegraphics[height=4cm]{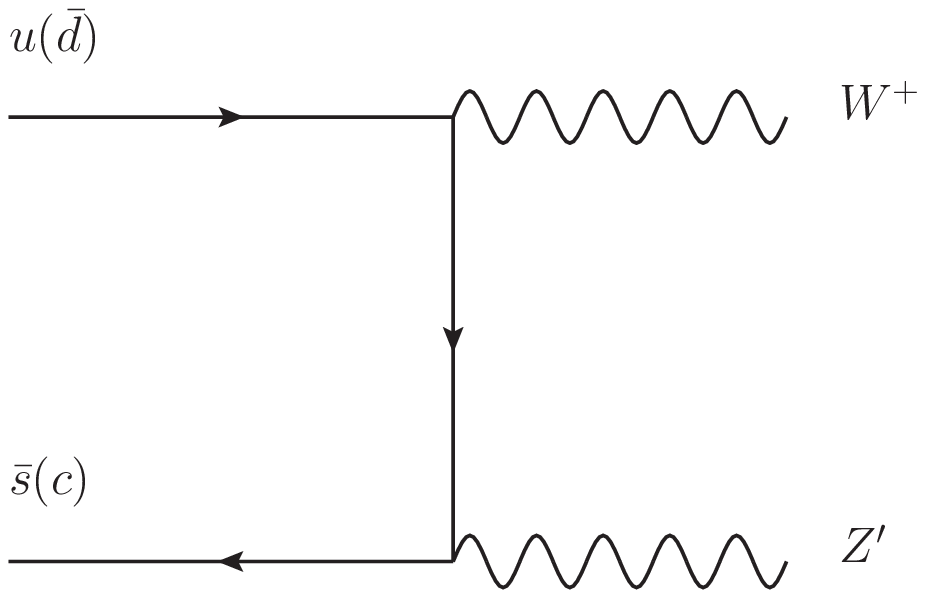}
\includegraphics[height=4cm]{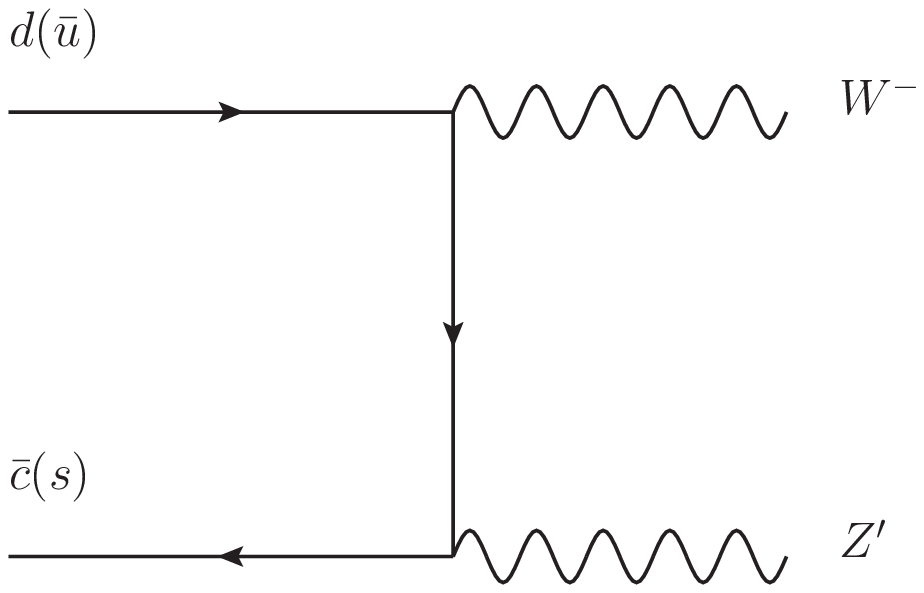}
\caption{\label{fig:Feyn:us}
The Feynman diagrams for $u \bar{s} \to W Z' \to W j j$ process.
}
\end{figure}

It is very likely that this CDF $Wjj$
anomaly is closely related 
with the SM $W$ boson.
We notice that the SM $W$ is the only gauge boson
which mediates flavor-changing current at tree level.
Based on these observations,
we propose a rather bold new physics model of $Z'$:
the $Z'$ couples only to the second generation quarks.
We denote this by $\zsc$
for the discrimination from the usual leptophobic $Z'$ 
with universal couplings to all generation quarks.
This $\zsc$ model definitely
introduces new flavor violation.
Recently the possibility of new flavor violation 
was shown in the CDF observation of
a large forward-backward asymmetry in $t\bar{t}$
production $A^{t\bar{t}}_{\rm FB}$ \cite{CDF:Afb:tt}
as well as the D0 measurement of 
the like-sign dimuon charge asymmetry \cite{D0:asl}.
$B$ physics also constrains 
quite significantly a flavor symmetric model
which can explain the CDF $Wjj$ anomaly
as well as $A^{t\bar{t}}_{\rm FB}$ \cite{B}.
One of the model classes to explain this new flavor violation
is to introduce the flavor violating couplings
of the SM quarks to new gauge bosons \cite{flavor:violation}.

Phenomenological consequences of our model are as follows.
First the $s$-channel single production 
of $\zsc$ at $p\pb$ collisions is
only through sea quarks, which
is suppressed by small parton distribution function (PDF) 
of sea quarks.
Second all the associated productions of $W \zsc$, $\gamma \zsc$,
and $Z\zsc$ 
through flavor-conserving 
channels
are from the second generation quarks:
small sea quark PDF effects
suppress this type of associated production.
Third the CKM quark mixing allows
additional flavor-changing contribution
to $W\zsc$ 
production from, \textit{e.g.}, $u\bar{s}$,
as depicted in Fig.~\ref{fig:Feyn:us}.
Small Cabbibo angle at the $u\bar{s}$ vertex
can be compensated by large valence quark PDF component.
This is the key motivation of our model to explain
the CDF $Wjj$ anomaly.
Finally the absence of $\zsc$-$b$-$\bar{b}$ coupling
leads to no effect on the search 
for $p \bar{p} \to WH \to \ell \nu b \bar{b}$ at the Tevatron.

Flavor-dependent $Z'$ models have been studied 
in the framework of the flavor symmetry 
like the Froggatt-Nielson mechanism \cite{fn}
with various motivations
like neutrino masses and fermion mass problems. 
Here we do not consider the underlying theory
but adopt a bottom-up approach.
From phenomenological perspectives discussed above,
we introduce the effective Lagrangian of $\zsc$ as
\be
\label{eq:Lg}
{\cal L}_{NC} = \gcs {\zsc^\mu} 
   \left( \bar{c}_L \gamma_\mu c_L + \bar{s}_L \gamma_\mu s_L \right).
\ee
We have assumed that $\zsc$ has the couplings only to
the left-handed second generation quarks.
This left-handed coupling maximizes the $W \zsc$
production while minimizing the $\gm\zsc$ production \cite{Wells}.

\begin{figure}[t!]
\centering
\includegraphics[height=4cm]{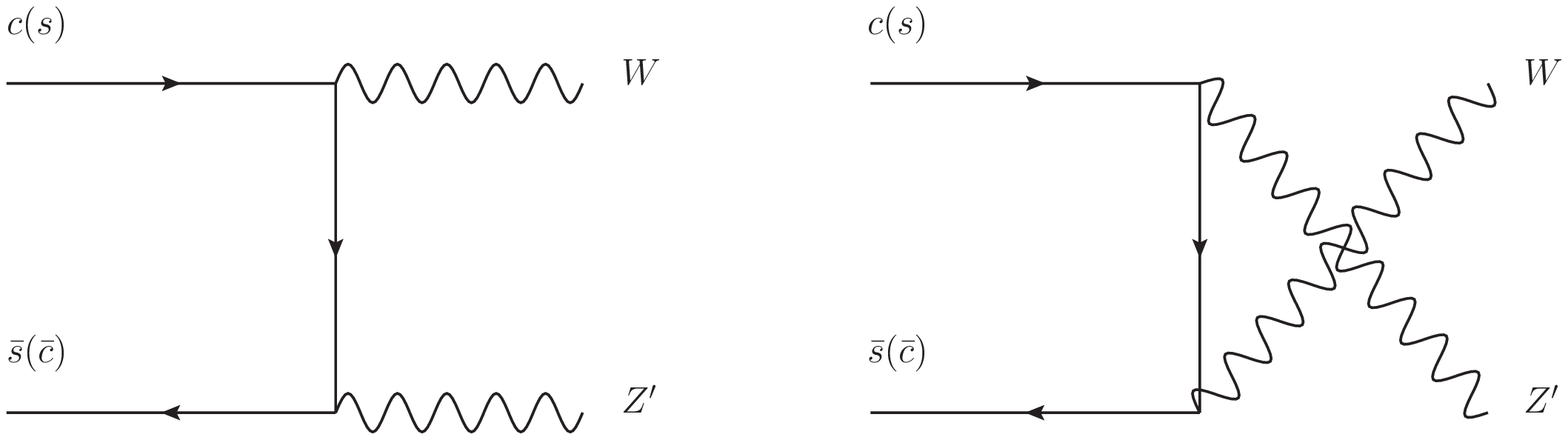}
\caption{\label{fig:Feyn:cs}
The Feynman diagrams for $c \bar{s} \to W Z' \to W j j$ process.
}
\end{figure}

The cross section is given by
\be
\sigma (p \bar{p} \to W^\pm \zsc)
= \int dx_a dx_b f_{q/p}(x_a) f_{q'/\pb}(x_b) 
\sum_{q,q'} \hat{\sigma}(q q' \to W^\pm \zsc),
\ee
where the parton level cross section for $W^+ \zsc$ production is
\be
\sum_{q,q'} \hat{\sigma}(q q' \to W^+ \zsc) =
 \hat{\sigma}(u \bar{s} \to W^+ \zsc) +
 \hat{\sigma}(\bar{d} {c} \to W^+ \zsc) +
 \hat{\sigma}(c \bar{s} \to W^+ \zsc).
\ee
Similar processes can be written for $W^- \zsc$ production.
The production channels of $u \bar{s}$ and $\bar{d} {c}$,
as illustrated in Fig.~\ref{fig:Feyn:us},
is mediated by one Feynman diagram.
It has the suppresstion of the quark mixing $|V_{us}|^2$ and 
the PDF of one sea quark.
Flavor-conserving channel of
$c \bar{s}$ as in Fig.~\ref{fig:Feyn:cs} 
has no quark mixing suppression, but additional sea quark
PDF suppression.

In the Tevatron $p\pb$ collisions at
$\sqrt{s}=1.96\tev$,
we have
\be
\sigma (p \bar{p} \to W^\pm \zsc) \simeq
K \left( \gcs \right)^2 \times 0.3 \hbox{ pb},
\hbox{ for } M_{\zsc} =145 \gev.
\ee
Note that Br$(\zsc \to j j) = 100$ \% in this model.
We have used the PDF of MRST \cite{Martin:2009iq}.
With the $K$-factor of $K=1.34$ \cite{K}, 
the observed $WZ'$ cross section of about 4 pb 
is explained by $\gcs \approx 3.2$.
This value of $\gcs$ marginally satisfies 
the perturbativity condition of
$\gcs^2 < 4 \pi$.
This large gauge coupling is attributed to
small PDF of a sea quark as well as the small quark mixing angle.

\begin{figure}[t!]
\centering
\includegraphics[height=8cm]{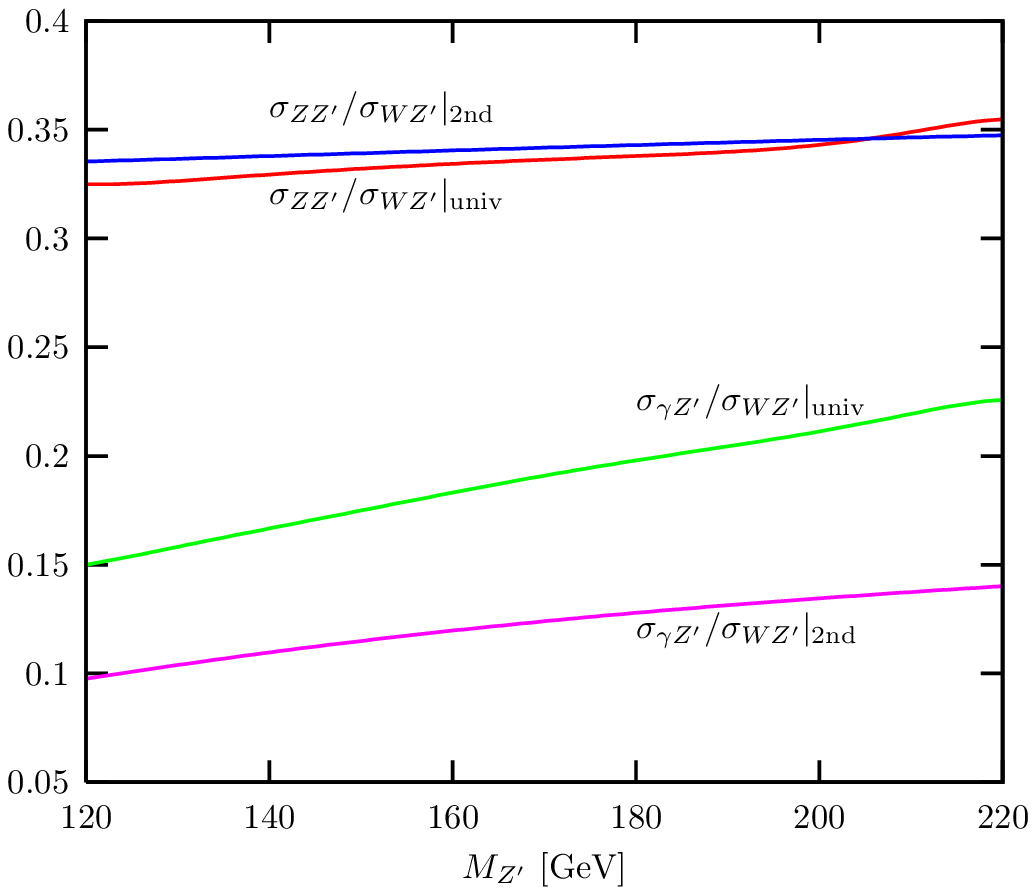}
\caption{\label{fig:ratio}
As functions of the new resonance mass,
the ratios of $\sigma(p\pb \to \gamma Z')/\sigma(p \pb \to W Z')$
and $\sigma(p\pb \to Z Z')/\sigma(p \pb \to W Z')$
in our $\zsc$ model and 
in the universally baryonic $Z'$ model \cite{kingman},
at the Tevatron $(\sqrt{s}=1.96\tev)$.
On the final state photon, we have imposed the acceptance of
$p_T^\gm > 50\gev$ and $|\eta^\gm|<1.1$. 
}
\end{figure}

Despite unpleasant largeness of the new gauge coupling $\gcs$,
our model has attractive features of suppressed $\gm jj$
and $\ell\nu b\bar{b}$ signatures at the Tevatron.
In Fig.~\ref{fig:ratio}, 
we compare the ratios of 
$\sigma(p\pb \to \gamma Z')/\sigma(p \pb \to W Z')$
and $\sigma(p\pb \to Z Z')/\sigma(p \pb \to W Z')$
in our $\zsc$ model and 
in the universally baryonic $Z'$ model \cite{kingman}.
(Note that $\sigma(W Z'_{(cs)})$ is fixed to be $\sim 4$ pb.)
For the photon acceptance we set 
$p_T^\gm > 50\gev$ and $|\eta^\gm|<1.1$.
As shown in Fig.~\ref{fig:ratio},
$Z Z'$ production cross sections
with respect to the $WZ'$ production
are almost the same in the universally baryonic
$Z'$ model and our $\zsc$ model.
The  $ZZ'\to Z jj$ signal
is a less sensitive probe because of small leptonic
branching ratios of $Z$.
Instead $\gm Z'$ production in our model
is about half of that in the universal $Z'$ model.

These behaviors are understood as follows.
In the universal $Z'$ model,
the main production channels of $\gm \zsc$ are
through the valence quarks of $u\bar{u}$ and $d\bar{d}$.
In a proton, the valence component of up quark with electric 
charge $+2/3$ is
twice of that of down quark 
with electric charge $-1/3$,
while the sea component of the $c$ quark is almost the same
as that of the $s$ quark:
the photon production from the valence quarks
has larger coupling strength
than that from the sea quarks.
On the contrary,
the $Z$ coupling to the left-handed quarks,
which is relevant in the associated production with the
left-handed $\zsc$ couplings in Eq.~(\ref{eq:Lg}),
has similar magnitudes for the up-type and down-type quarks.

Considering the lack of statistically significant 
excess of $\gm jj$ yet,
the reduced $\gm \zsc$ production in our model
compared to the universal $Z'$ model
can be one attractive merit.
Nevertheless the reduction of $\gm\zsc$ is not that large:
large enough luminosicy at the Tevatron will 
probe this $\gm jj$ excess in the very near future.
At the LHC energy,  the sea quark contributions are larger.
We expect that this $\gm jj$ process is more sensitive probe 
at the LHC.

In order to explain the recent CDF
excess of the $Wjj$ events
in the dijet invariant-mass distribution
as well as the (possible) lack of the excesses in the
$\gm jj$ and $Z jj$ events,
we proposed a new $\zsc$ model where the new gauge boson
couples only to the second generation quarks.
The CKM quark mixing allows more associated production 
with a $W$ boson, while suppressing the single production of $\zsc$,
and
the associated production of $\gm \zsc$ and $Z\zsc$.
If the new gauge coupling is large enough,
marginally permitted in a perturbative theory,
this new model can explain the observed CDF $Wjj$ anomaly.
Vanishing coupling $\zsc$ with the $b$ quarks
protects this model from the constraint
of the CDF and D0 searches for $WH \to \ell\nu b \bar{b}$.

\acknowledgments
This work is supported by WCU program through the KOSEF funded
by the MEST (R31-2008-000-10057-0).
KYL is also supported by
the Basic Science Research Program 
through the National Research Foundation of Korea (NRF) 
funded by the Korean Ministry of
Education, Science and Technology (2010-0010916).
SC is supported  by the Basic Science Research Program 
through the NRF funded by the Korean Ministry of
Education, Science and Technology  (KRF-2008-359-C00011).

\def\PRD #1 #2 #3 {Phys. Rev. D {\bf#1},\ #2 (#3)}
\def\PRL #1 #2 #3 {Phys. Rev. Lett. {\bf#1},\ #2 (#3)}
\def\PLB #1 #2 #3 {Phys. Lett. B {\bf#1},\ #2 (#3)}
\def\NPB #1 #2 #3 {Nucl. Phys. {\bf B#1},\ #2 (#3)}
\def\ZPC #1 #2 #3 {Z. Phys. C {\bf#1},\ #2 (#3)}
\def\EPJ #1 #2 #3 {Euro. Phys. J. C {\bf#1},\ #2 (#3)}
\def\JHEP #1 #2 #3 {JHEP {\bf#1},\ #2 (#3)}
\def\IJMP #1 #2 #3 {Int. J. Mod. Phys. A {\bf#1},\ #2 (#3)}
\def\MPL #1 #2 #3 {Mod. Phys. Lett. A {\bf#1},\ #2 (#3)}
\def\PTP #1 #2 #3 {Prog. Theor. Phys. {\bf#1},\ #2 (#3)}
\def\PR #1 #2 #3 {Phys. Rep. {\bf#1},\ #2 (#3)}
\def\RMP #1 #2 #3 {Rev. Mod. Phys. {\bf#1},\ #2 (#3)}
\def\PRold #1 #2 #3 {Phys. Rev. {\bf#1},\ #2 (#3)}
\def\IBID #1 #2 #3 {{\it ibid.} {\bf#1},\ #2 (#3)}


\begin{thebibliography}{99}

\bibitem{cdf} T. Aaltonen {\it et al.}, (CDF Collaboration),
arXiv:1104.0699 [hep-ex].

\bibitem{kingman} K. Cheung and J. Song, arXiv:1104.1375 [hep-ph].

\bibitem{light} 
M. R. Buckley, D. Hooper, J. Kopp, and E. Neil, arXiv:1103.6035 [hep-ph].

\bibitem{zp1} F. Yu, arXiv:1104.0243 [hep-ph];
X.-P. Wang, Y.-K. Wang, B. Xiao, J. Xu, and S.-h Zhu,
arXiv:1104.1161 [hep-ph].

\bibitem{Anchordoqui}
  L.~A.~Anchordoqui, H.~Goldberg, X.~Huang, D.~Lust, T.~R.~Taylor,
   [arXiv:1104.2302 [hep-ph]];
D.~-W.~Jung, P.~Ko, J.~S.~Lee,
 [arXiv:1104.4443 [hep-ph]];
M.~Buckley, P.~Fileviez Perez, D.~Hooper, E.~Neil,
 [arXiv:1104.3145 [hep-ph]];
P.~Ko, Y.~Omura, C.~Yu,
 [arXiv:1104.4066 [hep-ph]].

\bibitem{susy} C. Kilic and S. Thomas, arXiv:1104.1002 [hep-ph];
R. Sato, S. Shirai, and K. Yonekura, arXiv:1104.2014 [hep-ph].

\bibitem{techni} E. J. Eichten, K. Lane, and A. Martin, 
arXiv:1104.0976 [hep-ph].

\bibitem{color} 
X.-P. Wang, Y.-K. Wang, B. Xiao, J. Xu, and S.-h. Zhu,
arXiv:1104.1917 [hep-ph];
B. A. Dobrescu and G. Z. Krnjaic, arXiv:1104.2893 [hep-ph].

\bibitem{flavor} A. E. Nelson, T. Okui, and T. S. Roy, 
arXiv:1104.2030 [hep-ph].


\bibitem{intrinsic} X.-G. He and B.-Q. Ma, arXiv:1104.1894 [hep-ph].

\bibitem{SM:explanation}
Z.~Sullivan, A.~Menon,
 [arXiv:1104.3790 [hep-ph]];
 T.~Plehn, M.~Takeuchi,
 [arXiv:1104.4087 [hep-ph]].

\bibitem{Wells}
S.~Jung, A.~Pierce, J.~D.~Wells,
  [arXiv:1104.3139 [hep-ph]].

\bibitem{Tevatron_dijet} T. Aaltonen {\it et al.}, (CDF Collaboration),
\PRD 79 112002 2009 .

\bibitem{UA2} J. Alitti {\it et al.}, (UA2 collaboration),
\ZPC 49 17 1991 .

\bibitem{gamjj} CDF collaboration, CDF Note 10355.

\bibitem{CDF:Afb:tt}
T.~Aaltonen {\it et al.} [ CDF Collaboration ],
  Phys.\ Rev.\ Lett.\  {\bf 101}, 202001 (2008);
 T.~Aaltonen {\it et al.} [ CDF Collaboration ],
  [arXiv:1101.0034 [hep-ex]].

\bibitem{D0:asl}
V.~M.~Abazov {\it et al.} [ D0 Collaboration ],
  Phys.\ Rev.\  {\bf D82}, 032001 (2010).

\bibitem{B}
A.~E.~Nelson, T.~Okui, T.~S.~Roy,
  [arXiv:1104.2030 [hep-ph]];
 G.~Zhu,
  [arXiv:1104.3227 [hep-ph]].

\bibitem{flavor:violation}
S.~Jung, H.~Murayama, A.~Pierce, J.~D.~Wells,
  Phys.\ Rev.\  {\bf D81}, 015004 (2010);
V.~Barger, W.~-Y.~Keung, C.~-T.~Yu,
  Phys.\ Rev.\  {\bf D81}, 113009 (2010).
  [arXiv:1002.1048 [hep-ph]].
Q.~-H.~Cao, D.~McKeen, J.~L.~Rosner, G.~Shaughnessy, C.~E.~M.~Wagner,
  Phys.\ Rev.\  {\bf D81}, 114004 (2010).
  [arXiv:1003.3461 [hep-ph]].
J.~Shu, T.~M.~P.~Tait, K.~Wang,
  Phys.\ Rev.\  {\bf D81}, 034012 (2010);
K.~Cheung, T.~-C.~Yuan,
  Phys.\ Rev.\  {\bf D83}, 074006 (2011);
J.~Shelton, K.~M.~Zurek,
  [arXiv:1101.5392 [hep-ph]].



\bibitem{fn} C. D. Froggatt and H. B. Nielsen, \NPB 147 277 1979 . 

\bibitem{Martin:2009iq}
  A.~D.~Martin, W.~J.~Stirling, R.~S.~Thorne, G.~Watt,
  Eur.\ Phys.\ J.\  {\bf C63}, 189-285 (2009).
  [arXiv:0901.0002 [hep-ph]].

\bibitem{K}
U.~Baur, E.~L.~Berger,
  Phys.\ Rev.\  {\bf D47}, 4889-4904 (1993);
U.~Baur, T.~Han, J.~Ohnemus,
  Phys.\ Rev.\  {\bf D57}, 2823-2836 (1998);
J.~Huston,
  PoS {\bf RADCOR2009}, 079 (2010).



\end{thebibliography}
\end{document}